\documentclass[aps,prb,preprint,superscriptaddress,longbibliography]{revtex4-2}
\usepackage{graphicx}
\usepackage{braket}
\usepackage{amsmath}
\usepackage{amssymb}
\usepackage{bm}
\usepackage{color}
\begin{document}


\title{
Spontaneous frequency shift and phase delay of coupled terahertz radiations mediated by the Josephson plasmon in a cuprate superconductor
}


\author{Ryota Kobayashi}
\thanks{These authors contributed equally.}
\author{Ken Hayama}
\thanks{These authors contributed equally.}
\author{Shuma Fujita}
\affiliation{
Department of Electronic Science and Engineering, Kyoto University, Kyotodaigaku Katsura, Nishikyo, Kyoto 615-8510, Japan
}
\author{Manabu Tsujimoto}
\affiliation{
Research Center for Emerging Computing Technologies, National Institute of Advanced Industrial Science and Technology (AIST), Central2, 1-1-1 Umezono, Tsukuba, Ibaraki 305-8568, Japan.
}
\altaffiliation{
Faculty of Pure and Applied Sciences, University of Tsukuba, 1-1-1 Tennodai, Tsukuba, Ibaraki 305-8573, Japan
}

\author{Itsuhiro Kakeya}
\email{kakeya@kuee.kyoto-u.ac.jp}
\thanks{Corresponding author}
\affiliation{
Department of Electronic Science and Engineering, Kyoto University, Kyotodaigaku Katsura, Nishikyo, Kyoto 615-8510, Japan
}


\date{\today}

\begin{abstract}
We examine coupling interactions used to synchronize macroscopic Josephson oscillations induced in intrinsic Josephson junction (IJJ) mesa stacks made of a Bi2212 single crystal.
Synchronized radiations of terahertz electromagnetic (EM) waves are detected under common voltage and current bias operations of two connected mesas with close individual radiation frequencies, while uncoupled and bimodal radiations are frequently observed in two mesas with different individual radiation frequencies.
Detailed observations of the polarizations of the EM waves emitted when two mesas are biased in parallel or series allow us to reveal the coupling matrix components, which include ratios of synchronized IJJs in the mesas and  phase delay between the macroscopic Josephson oscillations.
A frequency evolution of the phase delay implies that the coupling between the Josephson oscillations is mediated by the small amplitude Josephson plasmon inside the superconducting substrate.
This finding stimulates systematic survey on polarization of EM wave emitted from synchronized multiple mesa devices in order to realize powerful terahertz emissions from superconductors.
\end{abstract}


\maketitle

\section{Introduction}

Various phenomena attributed to couplings between Josephson junctions have been observed in cuprate superconductor single crystals,
in which a strong modulation of order parameter along the $c$-axis attributed to the layered crystal structure exists~\cite{Kleiner:1994}.
In such systems, known as intrinsic Josephson junctions (IJJs), the excitation of a Josephson plasma wave along the $ab$-plane~\cite{Tachiki:1994,Kakeya1998b} induces macroscopic current oscillations on the crystal surface owing to the interplay with the alternating current (AC) Josephson effect and their synchronous oscillations accompanied by the inductive and the capacitive couplings among stacked IJJs~\cite{Sakai:1993,Koyama:1996,Machida1999,Mac:2004}, which combine to cause electromagnetic (EM) radiation into space~\cite{Machida:2000,Tachiki:2005,Bulaevskii:2007}.
Intensive research has been carried out on Josephson plasma emission (JPE) using mesa-structured devices formed from Bi$_2$Sr$_2$CaCu$_2$O$_{\rm 8+\delta}$ (Bi2212) single crystals~\cite{Ozyuzer:2007,Kadowaki2008,Minami2009,Hu2010,Wang2010,Yurgens2011,Kashiwagi2011,Koyama2011,Kakeya2012,Welp2013},
and further research on the possibility of a novel terahertz (THz) source have been reported during this decade~\cite{Li2012,Tsujimoto2014,Hao2015,Kakeya2016,Kashiwagi2017,Borodyanskyi2017,Elarabi2017,Elarabi2018,Kleiner2019,Benseman2019,Kuwano2020,Ono2020,Saito2021}.
For the implementation of this JPE as a practical terahertz source, it is indispensable to manage the synchronous oscillations resulting in higher radiation power and controlled performances.
However, the microscopic interaction between thousands of IJJs included in the mesa-devices have not been fully revealed partly because of the lack of experimental results in a single-layer IJJ with the similar $ab$-plane geometry to existing JPE devices.

One of the straightforward strategies to increase the radiation intensity is to synchronize multiple mesas to emit radiation. As an initial attempt, when two mesas were connected in series,
the coupled emission intensity was found to be higher than the sum of the individual intensities\cite{Orita2010}. Next, for a device with up to three mesas connected in parallel, an intensity of 0.61 mW was achieved, which is roughly proportional to the square of the number of mesas to be synchronized, and it was concluded that the emission of the electric wave is a synchronous coherent radiation~\cite{Benseman2013a}.

Very recently, the authors proposed a method for analyzing the synchronous phenomena with bases as polarizations of singly biased radiations and a coupling matrix between the Josephson plasma waves excited in the mesas from precise observations of the polarization of the radiated EM wave~\cite{Tsujimoto2020,Hayama2020}. This method allows us to determine the contribution of each mesa to the synchronized emission under biasing multiple mesas.
It is crucial to compare the emission features for parallel and series connections of a pair of mesas because the coupling matrix can be modified with the identical bases. However, there has been no report of investigations on the differences between them in an identical device.
Regarding a radiating mesa in which a thousand stacked IJJs oscillate coherently as a macroscopic Josephson junction~\cite{Lin2014}, the common voltage bias (CVB) produced by the parallel connection should induce synchronous oscillations between the macroscopic Josephson junctions, whereas the common current bias (CCB) produced by the series connection give insights for the interplay between a few stacked IJJs~\cite{Ota2009,Kitano2016,Nomura2015,Nomura2019}.

In this paper, we discuss the difference in inter-mesa coupling when two of three mesas are operated either by CVB or CCB modes, based on the polarization observation of EM waves emitted from mesas in individual and coupled emission states.
A CVB mostly generates a coupled emission with unimodal frequency spectrum, whereas a CCB tends to generate an uncoupled emission with bimodal frequency spectrum.
The coupled emission state is a superposition of the individual emissions perturbed by the inter-mesa coupling. The coefficients of the linear combination with the bases of the individual oscillation states
, which are chosen necessarily,
compose the mesa interaction matrices, i.e., the states of the harmonically oscillating plasmons inside the superconducting substrate that are responsible for the interaction between the {\em nonlinear} Josephson plasma oscillations in the mesas~\cite{Lin2013b}.
A systematic survey of the interaction matrix will lead to the development of powerful THz sources by controlling the mesa-to-mesa synchronization and will stimulate emergence of superconducting devices that enable THz quantum communication.

\section{Experiment}
In this study, we discuss EM wave radiations from three mesas (B, C, and E) formed on a Bi2212 single crystal grown by the floating zone method using photolithography and argon ion milling~\cite{Kakeya2012,Kakeya2016}. The device is shown in Fig. \ref{IVE} (a). The basic properties, such as the mesa shape, $T_c$, and current--voltage characteristics of the three mesas, are summarized in the Supplementary Materials. Parallel and series connections can simultaneously energize a pair of mesa structures connected by a Bi2212 substrate single crystal with CVB and CCB modes, respectively.
For a CVB, the wires connected to the two mesas were short-circuited at the outside of the cryostat, and a voltage was applied between the electrode and the substrate crystal, so that the same voltage was applied to the connected mesas, as shown in the right panel of Fig. \ref{IVE} (a).  On the other hand for a CCB, a bias voltages was applied between the two mesas, so it was not ensured that the same voltage was applied to each mesa.
An EM wave emitted was detected by a silicon bolometer cooled by liquid helium~\cite{Kakeya2016}, and the intensity was evaluated by the bolometer output voltage and the radiation frequency measured using an FTIR spectrometer with a split mirror\cite{Eisele2007,Wang2010,Kakeya2013}. For the polarization evaluation, the Stokes polarization parameters (SPPs) were estimated from the transmitted intensity through a polarization analyzer combining a fixed linear polarizer and a rotating achromatic quarter wave plate (QWP)~\cite{Nagai2015} as described in Refs. \cite{Schaefer2007,Tsujimoto2020,Polarizaion_analyser}.

\section{Results and discussion}
\subsection{Current--voltage characteristics and emission intensity}
\begin{figure}
 \includegraphics[width=1.0\columnwidth]{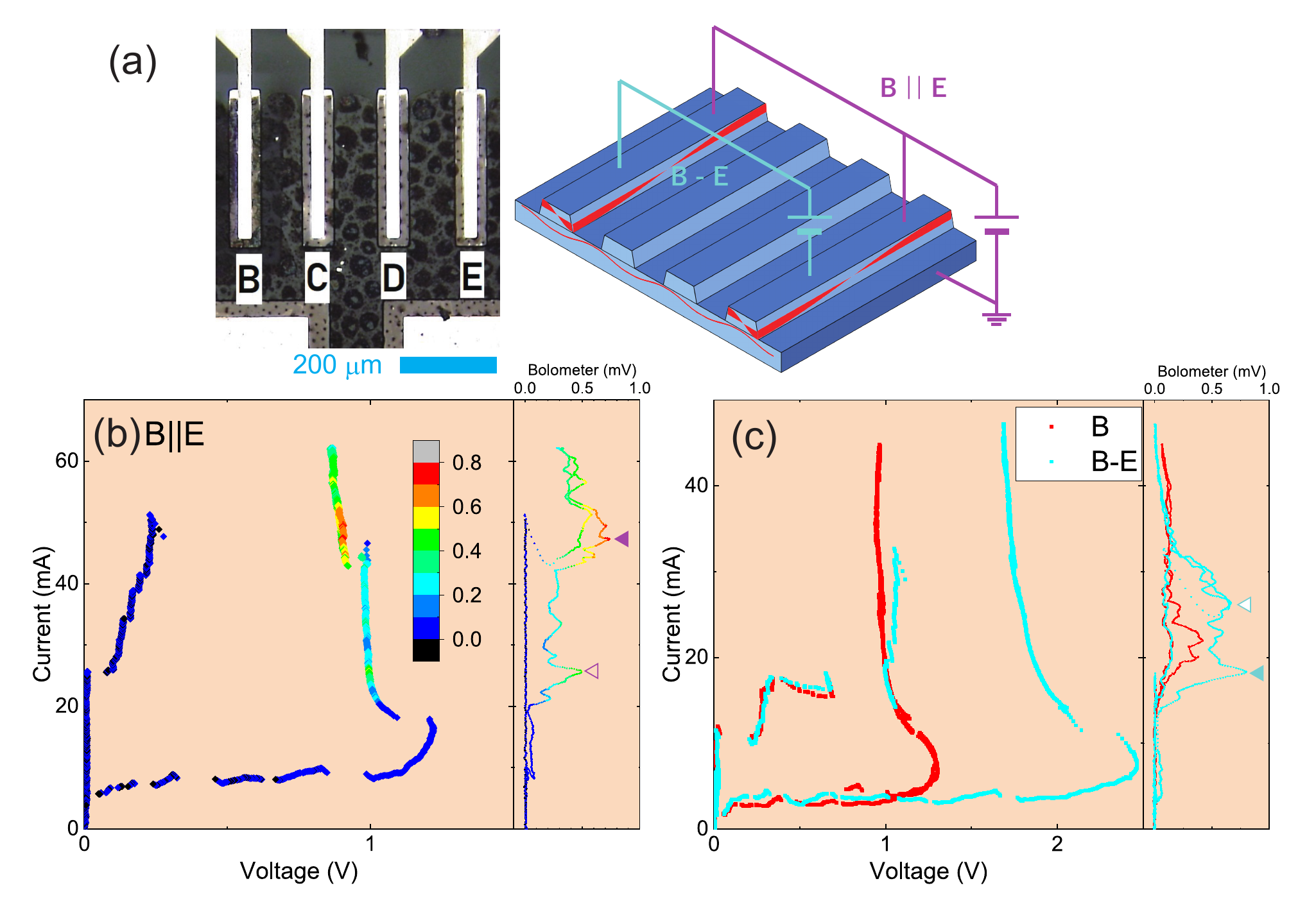}%
 \caption{(a) A microscope picture of the device (left) and a schematic drawing of the device and their connections for B $\parallel$ E (right,magenta) and B $-$ E (cyan). Properties of mesa D is not dealt in this paper.  Separations between mesas ($d$) are 87.5 (BC) and 412 (BE) $\mu$m.
 (b,c) Current-voltage characteristics and bolometer response of B$\parallel$E (b, color-coded), B (c, red), and B $-$ E (c, cyan).
 Solid and open triangles in the right panel of (b) and (c) point the first maximum and the second maximum in radiation intensity mentioned in the text, respectively.
 \label{IVE}}
 \end{figure}

Figure \ref{IVE} (b) shows the current--voltage--emission ($I-V-E$) characteristics of mesa B and E under CVB. This situation is represented by B$\parallel$E hereinafter. In comparison to data for individual mesas B and E (Fig. \ref{IVE}(c) and Fig. S1 in the Supplementary Materials), the current--voltage characteristics (IVC) for B$\parallel$E is elongated twice with respect to the current, and the emission intensity indicates more complicated multi-modal behavior as a function of the total bias current (Fig. \ref{IVE} (b) right panel).
First, the $I-V$ curve for B$\parallel$E with increasing bias is explained.
Below 25 mA, both B and E are superconducting and their bias currents are equal.
Between 25 and 50 mA, unequal currents flow in mesa B and E to keep CVB thus parts of IJJs of the mesas become resistive.
In the $E-I$ curve for  B$\parallel$E, two intensity maxima were observed: a broad maximum at $I=48$mA, and a rather sharp maximum at $I=26$ mA. The former current is slightly higher than the sum of the bias currents for the maximum emission intensity of both mesas B and E ($\simeq$ 22 mA, commonly). Here, the both mesas participate in the radiation and an additional current $\simeq$ 2 mA is supplied. The latter current approximately correspond to the sum of the bias current for the maximum emission intensity of either mesa B or E and the current ($\simeq$ 3 mA) for the same bias voltage region due to the multivalent IVC~\cite{Wang2010,Kakeya2012}. One mesa emits intensively with 22 mA, while the other does not under biasing at the same voltage.
As listed in Table \ref{table1}, the maximum detected intensity $P_{max}$ for B$\parallel$E is slightly smaller than the sum of $P_{max}$'s for B and E and the second maximum intensity for B$\parallel$E is very close to the $P_{max}$ for B or E.
Thus, we interpret that for B$\parallel$E both mesas B and E contribute to the emission above 40 mA while only either one contributes between 20 and 40 mA. The breakings of IV curve (at 40 and 18 mA) are attributed to the switchings of the internal synchronous phase dynamics of each mesas. This feature is consistent with a numerical calculation in a similar system consist of two IJJ stacks connected in parallel~\cite{Gross2015}.
In the subsequent discussion, therefore, spectra and polarization data above 40 mA for B$\parallel$E are employed for comparison between simultaneous and individual emissions.

A similar $I-V-E$ plot for the CCB of B and E (B$-$E) is shown in Fig. \ref{IVE}(c) together with a plot for B solo. The IVC for B$-$E below 1 V almost coincides with the IVC for B.
This is interpreted that only B contributes to the voltage and a large part of mesa E remains zero voltage because of the considerable difference in the distributions of the maximum Josephson currents of IJJs composing the mesas B and E seen in Fig. S1.
It was also found that the bolometer responses for B and B$-$E between 15 and 35 mA with increasing current are superposable. This clearly indicates that only B contributes to the emission power. In the same current range, with decreasing current, on the other hand, significant increase in bolometer response of B$-$E in comparison to the individual B was observed.
The maximum emission of B$-$E was attained at $I=18.4$ mA and the second maximum was found at $I=26.4$ mA (indicated by solid and open triangles, respectively).
These currents are approximately 4 mA lower and higher than the maximum radiating current of the individuals B and E, respectively. The net mesa voltage for B$-$E at the maximum emission power was $V=1.89$ V, which was slightly lower than the sum of the bias voltages for the individual emissions.
These shifts in bias current and voltage have been evaluated theoretically~\cite{Lin2013b}.
It is also interesting to note that the bolometer detection in decreasing current of B$-$E (and B) is not only large but also highly fluctuating in comparison to that in increasing current. This suggests that the coherent radiation is excited in the decreasing current while partly incoherent or chaotic radiation \cite{Gulevich2019,Shahverdiev2021} may be excited in the increasing current.

\subsection{Spectral characteristics}
\begin{figure}
 \includegraphics[width=0.9\columnwidth]{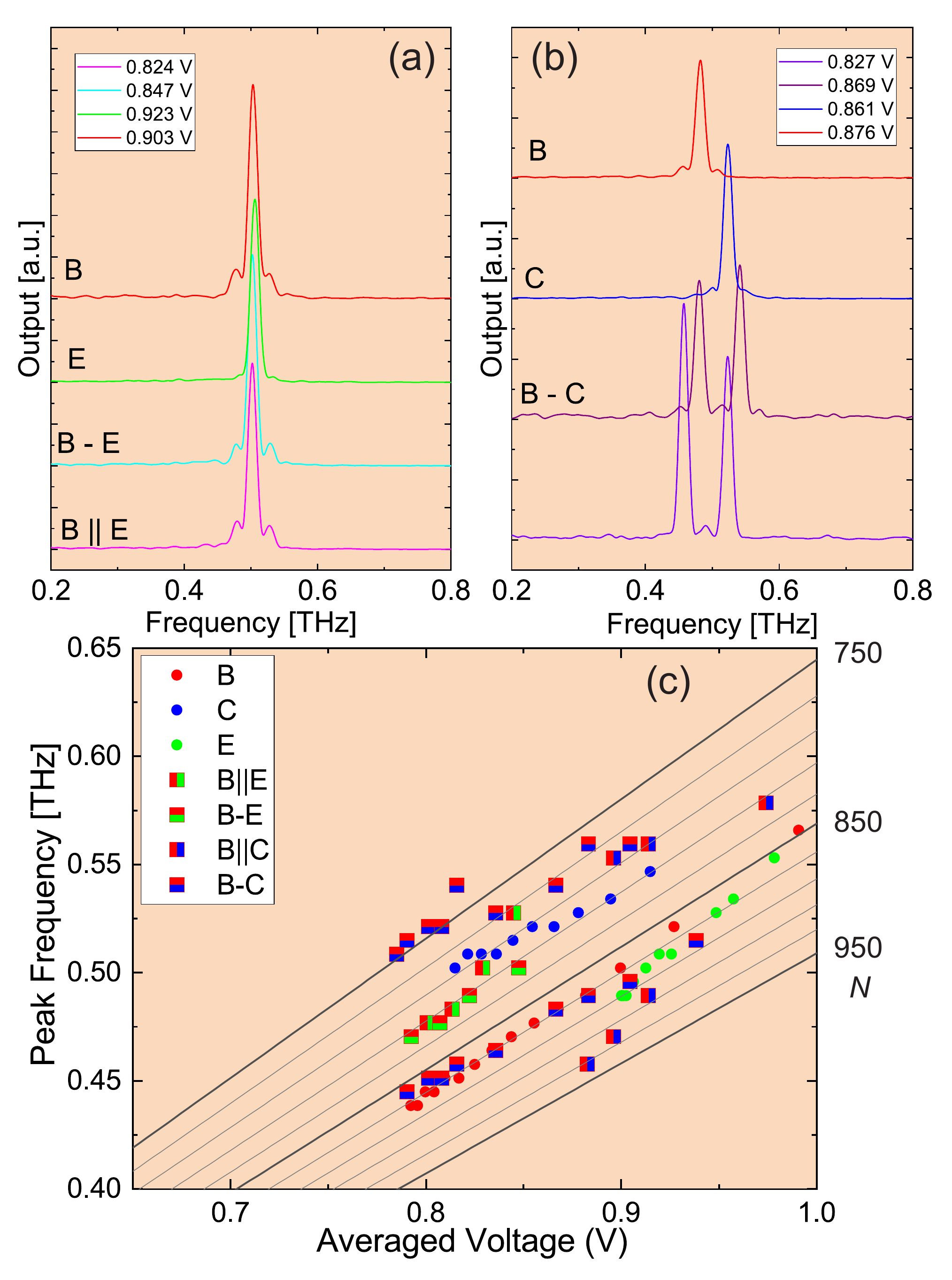}%
 \caption{FTIR spectra of coupled simultaneous radiations (a) and uncoupled simultaneous radiations (b) together with individual radiations at vicinity mesa voltages. In (c), peak frequencies of emission spectra are plotted as functions of averaged mesa voltage $V_{mesa}$: net voltage divided by 2 for series connections. Thick straight lines mean the ac Josephson relation with corresponding number of IJJs, $N$. Thin gray lines are drawn for $N$'s with a step of 20. \label{Spectra}
 All data are taken at bath temperature of 30 K.
 }
 \end{figure}

The radiation spectra of B, C, E, B$\parallel$E, B$-$E, and B$-$C are shown in Figs. \ref{Spectra}(a) and (b). Unimodal spectra were observed not only for individual operations but also for both CVB and CCB, suggesting that multiple mesas are coupled and radiating synchronously. The observed unimodal spectra were as follows: individual (B: 0.43--0.57, C: 0.50--0.55, E: 0.50--0.55 THz), B and E (CVB: 0.47--0.53, CCB: 0.46--0.5 THz), and C and E (CVB: 0.47--0.53, CCB: 0.48--0.61 THz). The frequencies of the emission spectra as functions of the averaged voltage applied to each mesa $V_{mesa}$ are shown in Fig. \ref{Spectra}(c). The linear fittings to the plots provide the number, $N$, of IJJs contributing to the radiations. For the BE pair, in which $N$'s for individual emissions of B and E are very close as $\sim 870$, and their effective $N$'s for CVB and CCB radiations are approximately 810, which is 8 \% smaller than those for B and E individuals.
Furthermore, radiation frequency region of B$\parallel$E and B$-$E is close to the overlapping frequency range of B and E, which have slightly shifted radiation frequency range due to the slight difference in mesa width.
Thus, the physics of EM radiation of the mesa B and E under simultaneous bias can be reworded as synchronous oscillations of two oscillators with slightly different eigenfrequencies with a cost of small portion of oscillating Josephson junctions.
For simultaneous bias of B and C (CVB: 0.45--0.58, CCB: 0.43--0.55 THz), bimodal spectra were observed, as shown in Fig. \ref{Spectra}(b). At this time, the mesas are presumed to be radiating independently. In particular, in the combination of B and C, a unimodal spectrum is observed only at 0.45 THz for CVB, while bimodal spectra were found in the other CVBs and all of the CCBs. The $N$ for C solo is distributed between 780 and 810, which is more than 10 \% smaller than that for B and E individuals.

The result that CVB operations tend to be coupled is understood as follows. For a CVB, an equal voltage is applied to the two mesas. Although it is not promised to apply an equal voltage is to the stacked IJJs composing a mesa structure, the inductive coupling among the IJJs\cite{Sakai:1993} expanding the whole mesa-stack results in an equal voltage being applied because of the Josephson oscillation of an identical frequency. Consequently, an equal voltage is applied to all contributing IJJs, and then Josephson plasma waves of a single frequency are excited in the CVB mesas. The BE and CE pairs, for which a unimodal spectrum is always observed, are presumed to have a relatively strong coupling between the mesas.

It is intriguing that the coupling in the BE pair with larger separation is relatively stronger than that in the combination of the BC pair resulting in synchronous oscillations. The reasons for this are discussed first.
The most important factor is the commensurability of the inter-mesa separation $d$ with respect to the mesa width $w$,
which is considered as a dominant factor in the emission frequency of a single mesa.
When we consider that the coupling is mediated by the Josephson plasmons, zero-voltage Josephson harmonic oscillations inside the superconducting substrate underneath the mesas, which has been observed as microwave absorption~\cite{Matsuda1995,Kakeya1998b}, the interference between the plasmons and the nonlinear Josephson oscillations excited in the mesas determines the coupling.
Here, using $w= 67 \mu$m (approximate average of widths of the mesas listed in Table SI of the Supplementary Materials), $d/w$ values of 1.3 (BC) and 6.1 (BE) are obtained.
For combinations with $d/w$ closer to an integer, the effective coupling attributed to the strength of the frequency entrainment of the nonlinear Josephson plasma oscillation is more pronounced. The second most important factor is the similarity in the device geometry, including electrodes: mesas B and E are mirror-symmetric while C is neither mirror- nor translational- symmetric with respect to B, and a large overlap in the emission frequency regions of individually biased cases are observed. In other words, for the cases that the device shapes are closer to being equivalent and the emission frequencies are closer, even weak perturbations may cause frequency entrainments, which result in synchronous radiations.
In summary, the BE pair provides coupled phase dynamics between two oscillators with \emph{close eigenfrequencies}.

\subsection{Polarization properties}
\begin{figure}
 \includegraphics[width=1.0\columnwidth]{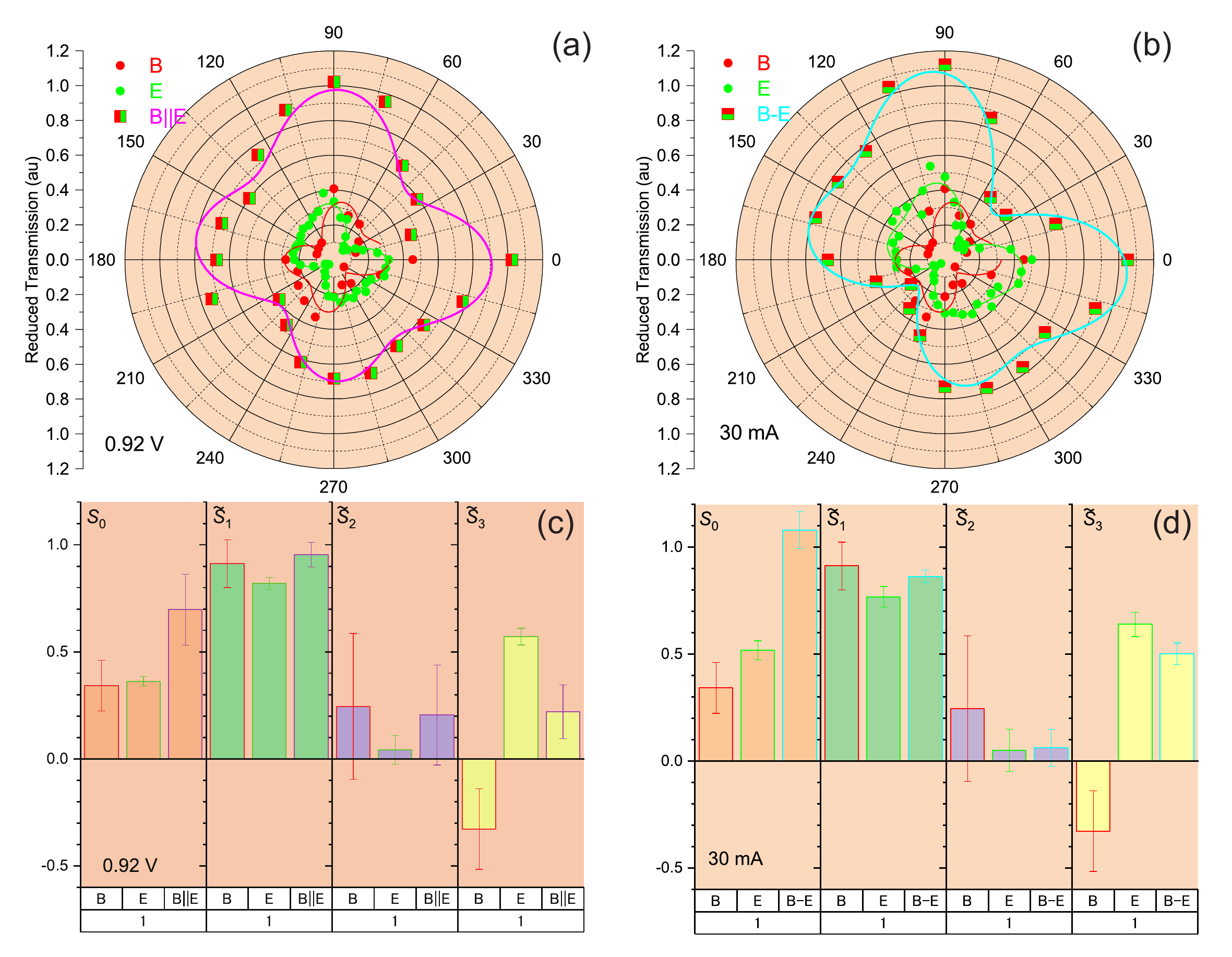}%
 \caption{(a,b) Polar plots of transmission intensity of B, E, B$\parallel$E (a), and B$-$E(b) through the polarization analyzer. Data at 0.92 V and 30 mA are plotted in (a) and (b), respectively. Symbols represent measured intensity and curves are least square fitting results of $ I_{corr}(\theta)$ given by Eqs. (\ref{eq:Polar}) and (\ref{eq:correction}) to derive SPPs.
(c,d) Derived SPPs from (a) and (b) are shown in (c) and (d), respectively.
$S_0$, $\tilde{S}_1$, and $\tilde{S}_2$ show no significant difference among B, E, B$\parallel$E andB$-$E.
$\tilde{S}_3$ representing helicity of EM vector has opposite sign between B and E. This is due to the mirror symmetry of the device shape between them.
 \label{Polar}
 }
 \end{figure}

Next, we discuss polarization on radiations from mesas B, E, and their combinations. Using the polarization analyzer described in the previous section, we obtain the transmitted intensity as a function of the QWP angle $\theta$ for B, E, B$-$E, and B$\parallel$E, as shown in Figs. \ref{Polar}(a) and (b).
Here, bias voltages to respective mesas are common as 0.92 V, for example. Based on these results, the Stokes polarization parameters SPPs $(S_0,S_1,S_2,S_3)=S_0(1, \tilde{S}_1, \tilde{S}_2, \tilde{S}_3)$ can be estimated by fitting
\begin{equation}
I\left(\theta\right)=\frac{1}{2}\left(S_0+S_1\cos^2{2\theta}+S_2\cos{2\theta}\sin{2\theta}+S_3\sin{2\theta}\right),
\label{eq:Polar}
\end{equation}
to the experimental data.
However, non-negligible two-fold asymmetry is found in the experimental data despite of the two-fold rotationally symmetric function of Eq. (\ref{eq:Polar}). This is attributed to an asymmetric loss of the QWP possibly due to deformations of its metallic plates.
See Sec. Siii in the Supplementary Materials.
In order to reduce this loss, we introduce a correction function $C(\theta)$ as
\begin{equation}
C(\theta)=1-\sqrt{a^2+b^2}+a \cos \theta+b \cos\theta,
\label{eq:correction}
\end{equation}
where $a$ and $b$ are constants determined by the optical system.
We employ $I_{corr}(\theta)=C(\theta)I(\theta)$ to fit the experimental data.
This operation corresponds to take a projection to original cross section of intensity from observed results.
The obtained SPPs for CVB and CCB are shown in Figs. \ref{Polar}(c) and (d), respectively. The values are listed in Table SII in Supplementary Materials.
Between mesa B and E, no significant difference is found for $S_0, \tilde{S}_1$, and $\tilde{S}_2$, whereas $\tilde{S}_3$ has the opposite sign, i.e., opposite helicity. This is attributed to the mirror symmetry of the device shapes including their electrodes.
Next, let us compare among individual and simultaneous bias cases.
$S_0$'s for both CVB and CCB cases considerably smaller than the square of the sum of $\sqrt{S_0}$'s for individual cases.
Significant difference in $\tilde{S}_3$ is found between CVB and CCB.
This is attributed to the phase delay to synchronize B and E.

\subsubsection{Linear combination method to describe a coupled emission}
Let us consider the coupled emission of the BE pair.
If we write electric field emitted from the individually biased mesas ${\text{B}}$ and ${\text{E}}$ as $\ket{{\text{B}}}=\bm{E}_{\text{B}}\exp[i(\omega_{\text{B}}t+\delta_{\text{B}} )]$  and $\ket{{\text{E}}}=\bm{E}_{\text{E}}\exp [i\left(\omega_{\text{E}}t+\delta_{\text{E}} \right)]$, respectively, where $\bm{E}_{\text{B}}$, $\omega_{\text{B}}$, and $\delta_{\text{B}}$ are the electric field vector, angular frequency, and the phase of an EM wave emitted from mesa ${\text{B}}$, respectively.
The simultaneous bias in the parallel connection can be written as follows:
\begin{equation}
\left|{\text{B}}\parallel {\text{E}}\right\rangle=\alpha\left|{\text{B}}\right\rangle+\beta\left|{\text{E}}\right\rangle,
\label{eq:LC-coupling}
\end{equation}
 where $\alpha$ and $\beta$ are complex.
 This is based on the idea that the observed coupled emission $\left|{\text{B}}\parallel {\text{E}}\right\rangle$ is a superposition of the EM waves emitted from mesas B and E under CVB, referred as $\ket{\text{B}^{\prime}}$ and $\ket{\text{E}^{\prime}}$, which are results of perturbation arising from the coupling interaction $\mathrm{V}_{{\text{B}\parallel \text{E}}}$ to $\ket{\text{B}}$ and $\ket{\text{E}}$, respectively.
 Thus,
\begin{equation}
\binom{\ket{\text{B}^{\prime}}}{\ket{\text{E}^{\prime}}}=\mathrm{V}_{\text{B}\parallel \text{E}} \binom{\ket{\text{B}}}{\ket{\text{E}}}=
    \begin{pmatrix}
    \alpha_1 & \beta_1 \\
    \alpha_2 & \beta_2 \\
    \end{pmatrix}
 \binom{\ket{\text{B}}}{\ket{\text{E}}},
\end{equation}
where $\alpha=\alpha_1+\alpha_2$ and $\beta=\beta_1+\beta_2$~\cite{Tsujimoto2020}.
Writing $\left|{\text{B}}\parallel {\text{E}}\right\rangle=\bm{E}_{{\text{B}}\parallel {\text{E}}}\exp [i\left(\omega_{{\text{B}}\parallel {\text{E}}}t+\delta_{{\text{B}}\parallel {\text{E}}}\right)]$,
 we obtain $\alpha\left(\omega_{\text{B}}\right)=|\alpha|\exp [i\left(\omega_{\text{B}}^{\prime} t+\delta_{\text{B}}^{\prime}\right)]$ and $\beta\left(\omega_{\text{E}}\right)=|\beta|\exp [i\left(\omega_{\text{E}}^{\prime} t+\delta_{\text{E}}^{\prime}\right)]$ with $\omega_{{\text{B}}\parallel {\text{E}}}=\omega_{\text{B}}+\omega_{\text{B}}^{\prime}=\omega_{\text{E}}+\omega_{\text{E}}^{\prime}$ and $\delta_{{\text{B}}\parallel {\text{E}}}=\delta_{\text{B}}+\delta_{\text{B}}^{\prime}=\delta_{\text{E}}+\delta_{\text{E}}^{\prime}$.
 Here, $\omega_{\text{B}}^{\prime}$  and $\delta_{\text{B}}^{\prime}$ are the frequencies and phase changes of mesa ${\text{B}}$ associated with synchronization, which indicates the magnitude of the inter-mesa perturbation.
 It is considered that $|\alpha|$ and $|\beta|$ correspond to the ratios of the number of IJJs contributing to for the coupled emission ($N_{\text{B}}^{\prime}$ and $N_{\text{E}}^{\prime}$) with respect to those for the individual emissions ($N_{\text{B}}$ and $N_{\text{E}}$); i.e. $|\alpha|=N_{\text{B}}^{\prime}/N_{\text{B}}$ and etc.
$\delta_{\gamma}= \arg\left(\beta/\alpha\right)=\delta_{\text{E}}^{\prime}-\delta_{\text{B}}^{\prime}$ is the phase difference of the coupling perturbation for mesa ${\text{B}}$ and ${\text{E}}$, which may related to the separation between the two mesas and the wavelength of the simultaneous oscillations of the two mesas.
In the following, we derive
\begin{equation}
({\alpha},{\beta})=\ket{\text{BE}}\binom{\ket{\text{B}}}{\ket{\text{E}}}^{-1},
\label{eq:alpha-beta}
\end{equation}
where $\ket{\text{BE}}$ is either $\ket{\text{B}\parallel {\text{E}}}$ or $\ket{\text{B}- \text{E}}$, and the validity of determining $\alpha$ and $\beta$ is discussed with fixing bases as specific polarizations of individual emissions in the following section.

\subsubsection{Fixed bias analysis results and discussion}
\label{sec:fixed_bias}

First, let us discuss synchronization between two mesas in CVB mode.
Since an equal voltage is applied for both mesas in the CVB mode, two sets of SPPs of $\ket{\text{B}}$, $\ket{\text{E}}$, and $\ket{\text{B}\parallel \text{E}}$ at two constant voltages are employed and compare their coefficients $\alpha$ and $\beta$ derived by Eq. (\ref{eq:alpha-beta}). Here, we assume that the oscillation frequency of the mesa does not change accompanied by the synchronization, i.e., $\omega_\text{B}^{\prime}=\omega_\text{E}^{\prime}=0$.
The coefficients for $V_{mesa}=$ 0.92 and 0.97 V are listed in Table \ref{table_2}.
The set of $(\alpha, \beta)$ at 0.92 V shows that the most of IJJs in both mesas B and E contribute to the radiation with 20 degrees phase difference. Here, the Josephson oscillation in the mesas B and E are apparently synchronized.
On the other hand at $V_{mesa}=$0.97 V, the obtained coefficients implies that only a small portion of IJJs composing the mesa B participates in the radiation for B$\parallel$E. Here, the phase difference $\delta_{\gamma}$ does not make sense. This would be attributed larger deviation in radiation frequency of B$\parallel$E from individual B.
As an example of CCB mode, we discuss $I_{mesa}=$ 30 mA, shown in Figs. \ref{Polar} (b) and (d). The obtained $|\alpha|$ is less than 0.3, suggesting that only 30 \% of IJJs of the mesa B participates in the synchronization. This seems to be consistent with the spectroscopic analysis that the CCB mode tends to show a bimodal FTIR spectrum.

In our previous publications~\cite{Tsujimoto2020,Hayama2020}, the bases used for the linear combination analysis were regarded as the polarizations with the intensity maximum of individual bias operations. A polarization of parallelly connected mesas was composed from two polarizations of single mesas at biases determined by radiation properties of respectively specified conditions. This procedure is considered to be the same as previous works for synchronous radiations of multiple mesas, where intensities and spectra at biases selected by their authors with implicit aspects are compared to discuss correspondences between individual and simultaneous bias operations\cite{Orita2010,Benseman2013a}. However, the requirement of common voltage or current either for parallel or series connection has not been necessarily taken into considerations.

\begin{figure}
\includegraphics[width=0.9\columnwidth]{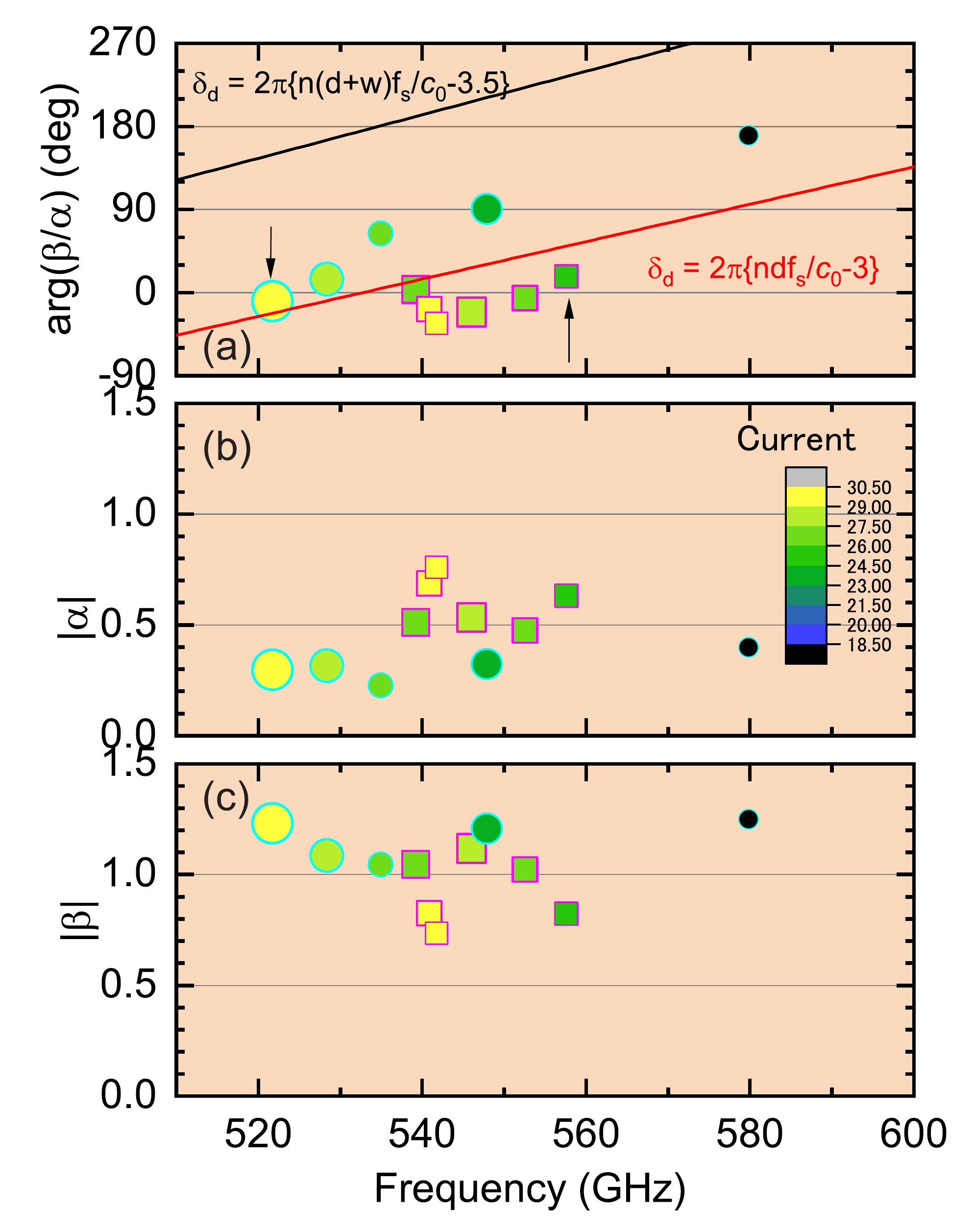}%
 \caption{
Estimated phase delay $\delta_{\gamma}$ (a), $|\alpha|$ (b), and $|\beta|$ (c) between the synchronized mesas as a function of frequency for CVB (square) and CCB (circle) operations.
Symbol size represents scaled $S_0$ and colors represent averaged bias current $I_{mesa}$ for the synchronized radiation.
Solid lines are phase difference corresponding to the mesa separation $d$ and mesa pitch $d+w$ of traveling waves inside a media with a refractive index $n=4.1$.
Arrows in (a) point data without frequency shift between individual and synchronized radiations, which are listed at rows (a) and (c) of Table \ref{table_2}.
 \label{fig:delta-f}
 }
 \end{figure}

Our analysis relies on three specific measured polarizations: a coupled emission with respect to two individual emissions. Considering the known facts on JPE, i.e., that emission intensity and polarization are sensitive to the temperature distribution of the device~\cite{Tsujimoto2014}, it is natural to presume that the temperature distribution of the simultaneous emission is different from the individual emissions. This discrepancy may result in deviations of $|\alpha|$ and $|\beta|$ from unity.
Let us apply these bases to the neighboring synchronous radiations to find frequency evolutions of the phase shift $\delta_{\gamma}$. Here, the frequency shift due to the mesa interaction $\omega^{\prime}$ is explicit.
Figure \ref{fig:delta-f} shows $\delta_{\gamma}$, $|\alpha|$, and $|\beta|$ for CVB radiations with bases at 0.92 V used in the previous paragraphs. Obtained parameters for CCB radiations with bases at 30 mA is also plotted.
It is found that the phase shift varies linearly with increasing frequency as the general trend common for CVB and CCB. The expected phase shift from the separation between the mesas $\delta_d=2\pi{ndf_s}/{c}_{0}$ is drawn with a solid line.
Noted that $\delta$ obtained with using 0.97 V bases shows no consistent trend.
For CCB, the evolution also follows to $\delta_d$ but the phase offset is different from CVB. The common slope in the frequency evolution of $\delta_{\gamma}$ close to $\delta_d$ determined by the refractive index of the superconducting substrate $n \simeq 4.1$ suggests that the interaction between the mesas are mediated not by electromagnetic waves in the space but by plasmons inside the Bi2212 single crystal.
An expected phase delay in accordance of the {\em pitch} of the mesas $\delta_{d+w}=2\pi{n(d+w)f_s}/{c}_{0}$ is also plotted in Fig. \ref{fig:delta-f} (a).
The agreement to the experimental $\delta_{\gamma}$ looks less likely than the case of $\delta_d$, however further analyses on synchronized radiations among closer mesas (smaller $d/w$; C and D for the present device) are required to judge the dominant distance factor ($d$ or $d+w$) for synchronized oscillation between mesas.

The systematic difference between CVB and CCB observed in the present study presents quantitative differences of media to promote the synchronous radiation; i.e., the Josephson plasmon excited the base crystal.
To attain highest radiation power  accompanied by fully controlled in-phase synchronization, $\delta_{\gamma}$ holds an essential relevance for CVB, and the number of synchronously oscillating IJJs within a mesa is another relevance for CCB.
In Ref. \cite{Lin2014}, it is reported that the phase delay between two synchronized IJJ stacks composing the same number of IJJs is zero.
The difference from the present study may be attributed to small difference in $N$, individual radiation frequency, inhomogeneous temperature rise, and so on.
These inevitable inequivalences of a real device can be compensated by an externally applied perturbation such as an external magnetic field or a focused laser irradiation.
We believe that the suggested polarimetry under the controlled perturbations and numerical analysis on Josephson junctions with modified coupling capacitances~\cite{Ottaviani2009} elucidate the mechanism for coherent phase dynamics of a Josephson junction array.
Furthermore, investigations on CVB and CCB systems with more mesas in combination with artificially-fabricated Josephson junction arrays~\cite{Galin2020} leads to multi-pixel two-dimensional JPE mesa array devices and a global coupling system of Josephson oscillators~\cite{Wiesenfeld1998} which represents synchronous phenomena within a mesa consisting of thousands IJJs.

\section{Summary}
Terahertz electromagnetic wave emission from two simultaneously biased mesas of Bi2212 IJJ stacks on a superconducting substrate was investigated by intensity, frequency, and polarization measurements as functions of bias voltage, combination of mesas, and parallel/series connections. Coupled and uncoupled emissions tended to be observed in parallel and series connections of the mesas, respectively. The frequencies of the coupled emissions were systematically varied.
Electromagnetic waves from pairs of coupled mesas can be represented by linear combinations of two individual emissions at shifted frequencies and phases, irrespective of parallel and series connections, rather than by simple superpositions of observed individual radiations modified by device temperature rise.
Further systematic investigations with parameters of inter-mesa separation and mesa geometries are required to fully assess the features of the perturbation matrix to control the inter-mesa synchronous phenomena and attain a high-power terahertz source from superconducting IJJ stacks.

\begin{acknowledgments}
This work was supported by the Japan Society for the Promotion of Science (JSPS) KAKENHI (Grant No. 20H02606) and
JSPS -- Centre national de la recherche scientifique (CNRS) Bilateral Program (Grant No. 120192908).
\end{acknowledgments}

\bibliography{library}

\clearpage

 \begin{table}
 \caption{
 Measured properties of coupled emissions for B$\parallel$E,  B$-$E, C$\parallel$E, and C$-$E.
 $P_{max}$ is the maximum detected radiation power represented bolometer response,  $I_{max}$ and $V_{max}$ are current and voltage yielding $P_{max}$, respectively.
 $N$ is number of IJJs contributing to the maximum radiation power averaged to a mesa for comparisons.
 \label{table1}
 }
 \begin{ruledtabular}
 \begin{tabular}{cccccc}
     &B  &C &E & B$\parallel$E  & B$-$E\\
\hline
 $P_{max}$ (mV)  & 0.418 & 0.215 & 0.468& 0.737 & 0.799 \\
 $I_{max}$ (mA) & 21.9 & 33.8 & 21.7& 47.3 & 18.3 \\
 $V_{max}$ (V) & 0.983 & 1.01 & 0.963& 0.903 & 1.899 \\
 $N$ & 867  & 796  & 879 & 798 & 815 \\
 \end{tabular}
 \end{ruledtabular}

 \end{table}


 \begin{table}
 \caption{
Obtained linear-coupling coefficients $\alpha, \beta$, and $\delta_{\gamma}=\arg(\beta/\alpha)$.
 \label{table_2}
 }
 \begin{ruledtabular}
 \begin{tabular}{lccccc}
& & Bias & $|\alpha|$ & $|\beta|$  & $\arg(\beta/\alpha)$ (deg.) \\
\hline
(a) & B$\parallel$E  &0.92 V& 0.632	&0.824&	17.4\\
(b) & B$\parallel$E &  0.97 V & 0.145	& 0.953&	306.1 \\
\hline
(c) & B$-$E  & 30 mA & 0.297&	1.23&	-9.02 \\
 \end{tabular}
 \end{ruledtabular}
 \end{table}

\clearpage

\end{document}